\documentclass[10pt,conference]{IEEEtran}

\usepackage{microtype}
\usepackage{graphicx}
\usepackage{booktabs}
\usepackage{amsmath,amssymb}
\usepackage{enumitem}
\usepackage{hyperref}
\usepackage{url}
\usepackage{xcolor}
\usepackage{listings}
\usepackage{algorithm}
\usepackage{algpseudocode}
\usepackage{float}

\hypersetup{
  colorlinks=true,
  linkcolor=blue,
  citecolor=blue,
  urlcolor=blue,
  hypertexnames=false
}

\title{Guardrails as Infrastructure: Policy-First Control for Tool-Orchestrated Workflows}
\author{
Akshey Sigdel \\
Independent Researcher \\
aksheysigdel@u.boisestate.edu \\
\and
Rista Baral \\
Independent Researcher \\
ristabaral@u.boisestate.edu \\
}
\date{}

\begin{document}
\maketitle

\vspace{-15ex}
\begin{abstract}
Tool-using automation systems, from scripts and CI bots to agentic assistants, fail in recurring patterns. Common failures include unsafe side effects, invalid arguments, uncontrolled retries, and leakage of sensitive outputs. Many mitigations are model-centric and prompt-dependent, so they are brittle and do not generalize to non-LLM callers. We present \emph{Policy-First Tooling}, a model-agnostic permission layer that mediates tool invocation through explicit constraints, risk-aware gating, recovery controls, and auditable explanations. The paper contributes a compact policy DSL, a runtime enforcement architecture with actionable rationale and fix hints, and a reproducible benchmark based on trace replay with controlled fault and misuse injection. In 225 controlled runs across five policy packs and three fault profiles, stricter packs improve violation prevention from 0.000 in P0 to 0.681 in P4, while task success drops from 0.356 to 0.067. Retry amplification decreases from 3.774 in P0 to 1.378 in P4, and leakage recall reaches 0.875 under injected secret outputs. These results make safety to utility trade-offs explicit and measurable.
\end{abstract}

\section{Introduction}
Tool-orchestrated workflows are now common in software engineering. CI pipelines run builds and tests, bots update dependencies, and assistants automate repository operations. These workflows share one critical property. Tools can create irreversible side effects. One unsafe or malformed invocation can delete files, leak secrets, trigger expensive retries, or corrupt state.

A growing body of work studies LLM-based tool use and reliability, including interleaved reasoning and acting, self-supervised tool selection, and program-aided execution \cite{yao2023react,schick2023toolformer,gao2023pal}. Recent agent frameworks and API-benchmarked systems show strong capability but still rely on reliable tool governance to manage side effects \cite{shen2023hugginggpt,patil2023gorilla,qin2023toollm}. Recent controlled studies in schema-first tool APIs and embodied behavior generation further motivate explicit operational constraints and measurable recovery behavior \cite{sigdel2026schemafirst,baral2025novel}. Model-centric guardrails remain insufficient when failures arise from weak contracts, missing invariants, lack of idempotency, and unbounded retries. Many production settings also require local CPU-only execution and model-agnostic safety controls.

\paragraph{Thesis.} A policy-first permission layer with explicit constraints, risk-aware gating, and recovery controls can significantly reduce tool misuse and unsafe side effects in developer automation while preserving developer throughput and enabling graceful recovery under realistic failures and budget constraints. This claim can be evaluated through reproducible and LLM-independent benchmarks based on fault injection and trace replay.

\paragraph{Research Objective.}
Design, implement, and rigorously evaluate a policy-first permission layer for tool invocation across filesystem, network, shell, and API tools. The objective is to prevent misuse and unsafe side effects while preserving developer throughput and enabling graceful recovery under realistic failures and budget constraints.

\paragraph{Thesis Focus (Core).}
The work centers on constrained execution with tool allow and deny rules, argument constraints, rate limits, and approvals for high-risk actions. It emphasizes safety by construction through blocking destructive deletes, writes outside the workspace, and secret leakage. It addresses reliability through retry policy, backoff, circuit breakers, and idempotency. It also enforces auditability by logging decisions with ``why blocked / how to fix'' explanations.

\paragraph{Non-goals.}
We do not attempt to ``make the LLM smarter.'' The permission layer must operate for any caller, including buggy scripts, small local models, or human-invoked tooling.

\paragraph{Scope and Practical Constraints.}
Evaluation is CPU-only and LLM-independent, and it relies on trace replay, simulated tools, and deterministic fault and misuse injection.

\paragraph{Contributions.}
We contribute a model-agnostic permission layer for tool invocation with constraints, invariants, budgets, approvals, and redaction. We define a concise policy DSL with allow and deny rules, argument constraints, cross-field invariants, cost and rate budgets, approval gates, and output redaction. We implement a runtime Policy Enforcement Point that intercepts tool calls and records auditable rationale with actionable fix hints. We provide a reproducible benchmark suite with a trace schema, a fault catalog, a misuse taxonomy, and standardized metrics. We derive empirical insights on which policy primitives provide the strongest safety to utility trade-offs.

\paragraph{Research Questions.}
We study five questions. \textbf{RQ1 Misuse Prevention} asks how effectively the policy layer blocks unsafe or invalid actions relative to schema-only validation and ad hoc checks. \textbf{RQ2 Productivity Cost} measures overhead in latency, blocks, and approvals. \textbf{RQ3 Recovery} examines which policy-driven recovery controls maximize success under faults without cost amplification. \textbf{RQ4 Policy Design} identifies which primitives produce the strongest Pareto frontier. \textbf{RQ5 Explainability} tests whether policy explanations reduce time to fix after denials and improve developer trust.

\section{Background and Related Work}
This section positions policy-first tooling within four related areas. These areas include API contracts and schema validation, capability-based security and policy enforcement, reliability patterns in distributed systems, and guardrails for tool-using assistants.

\paragraph{Contracts and Validation.}
Schema validation, such as JSON Schema, catches type-level errors but rarely captures richer invariants and runtime budgets. This limitation is consistent with long-standing security policy theory, where enforceability depends on explicit reference monitoring and policy semantics beyond structural type checks \cite{schneider2000,saltzer1975}.

\paragraph{Policy Enforcement and Capabilities.}
Security research provides mature concepts including least privilege, capability tokens, and reference monitors. Classical confinement and capability systems demonstrate why ambient authority should be minimized and enforcement should be explicit at invocation boundaries \cite{lampson1974,hardy1988,watson2015cheri}. Role- and policy-based access control frameworks further motivate configurable, auditable decision points for operational systems \cite{sandhu1996,schneider2000}.

\paragraph{Reliability Patterns.}
Classic resilience mechanisms including timeouts, retries, jitter, and circuit breakers mitigate failure. Without budgets and idempotency, the same mechanisms can amplify side effects. This trade-off is aligned with tail-latency and production reliability findings in large-scale systems \cite{dean2013tail,beyer2016sre}.

\paragraph{Tool-Using Assistants.}
LLM-centric guardrails such as prompting, constrained generation, and constitutional constraints are complementary but they do not replace system-level enforcement \cite{yao2023react,schick2023toolformer,bai2022constitutional}. Recent agent systems that orchestrate many tools and APIs further motivate policy-first runtime mediation for safety, reliability, and auditability \cite{shen2023hugginggpt,patil2023gorilla,qin2023toollm}. Recent results from embodied behavior generation and schema-first API studies support the same conclusion that explicit policy boundaries and recovery controls are essential \cite{baral2025novel,sigdel2026schemafirst}.

\section{System and Threat Model}
\subsection{System Entities}
We model a tool-orchestrated workflow with five entities. A \textbf{Caller} $C$ issues tool calls and may be a script, bot, agent, or user interface. \textbf{Tools} $\mathcal{T}$ are side-effecting functions such as filesystem, shell, HTTP, and database operations. A \textbf{Policy Enforcement Point (PEP)} $E$ intercepts calls. A \textbf{Policy Decision Point (PDP)} $P$ evaluates policies over calls and context. A \textbf{Telemetry and Audit Log} $L$ records calls, decisions, outcomes, and explanations.

\subsection{Call and Decision Semantics}
A tool call is a tuple of
\[
  \langle \texttt{tool\_name}, \texttt{args}, \texttt{context}, \texttt{budget}, \texttt{metadata} \rangle
\]
where context can include workspace root, repository state, execution environment, and prior calls. The PDP returns one of four outcomes
\[
  d \in \{\texttt{ALLOW}, \texttt{DENY}, \texttt{REQUIRE\_APPROVAL}, \texttt{TRANSFORM}\}
\]
together with rationale, developer-facing fix hints, and optional transformations such as redaction.

\subsection{Trust Assumptions and Boundaries}
We focus on \textbf{accidental misuse} and \textbf{bounded adversarial inputs}, including untrusted file paths, log strings, and repository content. We assume tools are not malicious at implementation level. Tools may still fail, return malformed outputs, or expose sensitive strings in outputs.

\subsection{Misuse Taxonomy}
\begin{enumerate}[leftmargin=*]
  \item \textbf{Unsafe Side Effects}. Destructive operations such as delete or overwrite outside the workspace.
  \item \textbf{Invalid Invocation}. Missing required arguments, wrong types, or invalid enums and ranges.
  \item \textbf{Budget Violations}. Runaway retries, excessive calls, and expensive operations.
  \item \textbf{Data Leakage}. Secrets in outputs or unintended exfiltration through network tools.
  \item \textbf{State Corruption}. Partial writes and non-idempotent retries.
\end{enumerate}

\subsection{Out of Scope}
We do not target sophisticated adaptive adversaries with root access. The scope targets realistic engineering hazards and mitigations that can be enforced through policy.

\section{Policy Architecture and Enforcement}
\subsection{Policy Primitives}
The policy layer supports tool gating by tool, group, or capability. It supports argument constraints based on regex, range, enum, and path containment. It supports cross-field invariants through conditional rules and mutual exclusion. It supports budgets with call limits, retry limits, cost ceilings, and rate limits. Approval gates enable human review above a risk threshold. Redaction sanitizes outputs with taint-like propagation. Recovery controls include retry, backoff, jitter, circuit breakers, and fallbacks.

\subsection{Auditability and Developer Guidance}
Each decision includes a human-readable rationale and a fix hint, such as path under workspace root or retry budget exhausted with a recommendation to adjust budget or call rate. These explanations are logged with each call and support evaluation of time to fix and trust.

\subsection{Risk Scoring and Approval}
We define an additive risk score as
\[
  R(\texttt{call}) = w_t \cdot r(\texttt{tool}) + w_a \cdot r(\texttt{args}) + w_c \cdot r(\texttt{context})
\]
Approval is required when $R(\texttt{call}) \ge \tau$.

\subsection{Illustrative Policy DSL}
\begin{lstlisting}
policy "workspace_fs_safety" {
  tool_group: ["fs.read", "fs.write", "fs.delete"]
  allow_if: args.path starts_with context.workspace_root
  deny_if:  args.path matches "(^ROOT/|\\.\\./|~)"  // no root-absolute, no traversal, no home
  require_approval_if: tool == "fs.delete" and args.recursive == true
}

policy "shell_safety" {
  tool: "shell.exec"
  deny_if: args.cmd matches "(rm\\s+-rf\\s+ROOT|:(){:|:&};:|mkfs\\.)"
  budget: max_calls_per_minute = 10
}

policy "output_redaction" {
  on_output: true
  redact_patterns: ["AKIA[0-9A-Z]{16}", "-----BEGIN PRIVATE KEY-----"]
}
\end{lstlisting}

\subsection{Enforcement Control Flow}
The execution path is straightforward. The caller issues a tool call, the PEP intercepts it, and the PDP evaluates policy. The decision permits execution, denies execution, requests approval, or applies transformation. If execution proceeds, outputs are redacted as needed and the system records the audit log.

\subsection{PEP Decision Algorithm}
\begin{algorithm}[H]
\caption{Policy Enforcement for Tool Calls}
\label{alg:pep}
\begin{algorithmic}[1]
\Require tool\_call $c$, policy\_set $\Pi$, state $S$ (budgets, history)
\State $d \gets \textsc{Evaluate}(\Pi, c, S)$
\State \textsc{LogDecision}($c, d$)
\If{$d$ is \texttt{DENY}}
  \State \Return error(\texttt{policy\_denied})
\ElsIf{$d$ is \texttt{REQUIRE\_APPROVAL}}
  \State $a \gets$ \textsc{RequestApproval}($c, d$)
  \If{not $a$} \State \Return error(\texttt{approval\_rejected}) \EndIf
\EndIf
\State $c' \gets$ \textsc{ApplyTransforms}($c, d$) \Comment{e.g., sanitize args}
\State $o \gets$ \textsc{ExecuteTool}($c'$)
\State $o' \gets$ \textsc{RedactOutput}($o, \Pi, S$)
\State \textsc{UpdateState}($S, c', o'$)
\State \textsc{LogOutcome}($c', o'$)
\State \Return $o'$
\end{algorithmic}
\end{algorithm}

\subsection{Design Principles}
\begin{itemize}[leftmargin=*]
  \item \textbf{Model-agnostic}. Callers may be LLM-based or non-LLM.
  \item \textbf{Deterministic and auditable}. Identical inputs and policy produce identical decisions.
  \item \textbf{Low overhead}. Safe defaults, lightweight scoring, and fast validation.
  \item \textbf{Composable}. Multiple policies can be layered with configurable strictness.
\end{itemize}

\section{Benchmark Suite Specification}
\label{sec:benchmark}

\subsection{Overview}
The benchmark operationalizes the system model and policy architecture through \textbf{trace replay} and \textbf{fault and misuse injection}. It does not require an LLM. Callers can be deterministic scripts that emit tool calls from traces, heuristic planners, or optional local model baselines.

\subsection{Benchmark Components}
The benchmark includes a standardized tool registry with wrappers for filesystem, shell, HTTP, and database mock interfaces. It defines a canonical JSONL trace format for tool-call sequences and a matching decision log format. A misuse injector introduces unsafe arguments, invalid enums, and leak strings. A fault injector adds timeouts, rate limits, corrupt outputs, and partial writes. Policy packs span permissive through strict settings. A metrics harness computes safety, reliability, utility, and overhead.

\subsection{Tool Registry}
\paragraph{Minimum Tool Set (CPU-only).}
The minimum tool set includes filesystem tools \texttt{fs.read}, \texttt{fs.write}, and \texttt{fs.delete}. It includes \texttt{shell.exec} in a restricted sandbox and \texttt{http.get} in mocked or offline mode. It includes parsing tools \texttt{parse.json} and \texttt{parse.regex}. It also includes \texttt{kv.get} and \texttt{kv.put} for an idempotency ledger.

\paragraph{Side-effect Semantics.}
Each tool declares whether it is read-only, write, delete, network, or exec. This metadata is consumed by policy evaluation and risk scoring.

\subsection{Trace Format (JSONL Spec)}
Each line is a JSON object that represents one call. The trace is replayable and includes expected outcomes and annotations.

\begin{lstlisting}
{
  "trace_id": "task_fs_refactor_001",
  "step_id": 7,
  "ts": "2026-02-27T15:04:05Z",
  "tool": "fs.write",
  "args": { "path": "repo/src/config.yaml", "contents_ref": "blob:9f2a", "mode": "overwrite" },
  "context": {
    "workspace_root": "repo",
    "repo_state": "git:abc123",
    "env": "local_cpu"
  },
  "budget": { "max_calls": 40, "max_cost": 120, "deadline_ms": 60000 },
  "annotations": {
    "safety_class": "write_in_workspace",
    "expected": "success",
    "expected_side_effect": "file_modified"
  }
}
\end{lstlisting}

\paragraph{Required Fields.}
Required fields are \texttt{trace\_id}, \texttt{step\_id}, \texttt{tool}, \texttt{args}, and \texttt{context}.

\paragraph{Optional Fields.}
Optional fields include budget and annotations, plus expected output hashes and dependency links.

\subsection{Policy Decision Log Format}
The harness emits a separate JSONL decision log. Trace records and decision records share \texttt{trace\_id} and \texttt{step\_id} so that analyses can join both streams deterministically.

\begin{lstlisting}
{
  "trace_id": "task_fs_refactor_001",
  "step_id": 7,
  "decision": "REQUIRE_APPROVAL",
  "policy_ids": ["workspace_fs_safety"],
  "rationale": ["fs.write allowed only under workspace_root", "overwrite requires approval for *.yaml"],
  "fix_hint": "Use a path under workspace_root or request approval for overwrite.",
  "risk_score": 0.71,
  "transforms_applied": [],
  "overhead_ms": 2.3
}
\end{lstlisting}

\subsection{Misuse Taxonomy and Injection Specification}
Misuse injection perturbs traces to create controlled negative cases.

\paragraph{Misuse Types.}
Misuse types include path traversal by inserting \texttt{"../"} or absolute paths into filesystem calls. They include destructive deletes outside the workspace, command hazards in shell strings, and data exfiltration through \texttt{http.get} calls to unapproved domains or with secret-bearing query parameters. They include schema invalidity through wrong types, enums, or ranges, and secret-in-output injections for redaction tests.

\paragraph{Injection Config (YAML)}
The following YAML snippet specifies a deterministic misuse injection setup.
\begin{lstlisting}
misuse_injection:
  rate: 0.20                 # fraction of eligible calls mutated
  seed: 1337
  mutations:
    - type: path_traversal
      tools: ["fs.read", "fs.write", "fs.delete"]
      patterns: ["../", "etc/passwd", "home/user/.ssh/id_rsa"]
    - type: dangerous_shell
      tools: ["shell.exec"]
      patterns: ["rm -rf ROOT", "mkfs.ext4", ":(){:|:&};:"]
    - type: schema_invalid
      tools: ["http.get"]
      fields: ["timeout_ms"]
      values: [-1, "fast", 999999999]
    - type: secret_in_output
      tools: ["http.get", "shell.exec"]
      patterns: ["AKIA................", "-----BEGIN PRIVATE KEY-----"]
\end{lstlisting}

\subsection{Fault Injection Catalog}
Fault injection models realistic tool failures.

\paragraph{Fault Types.}
Fault types include timeouts when a tool exceeds \texttt{timeout\_ms}. They include transient errors that fail and then succeed with probability $p$ on retry. They include rate limits that emulate 429 windows. They include corrupt outputs such as invalid JSON or truncated text. They include partial writes that succeed only in part and then return an error. They include non-determinism where outputs differ across repeated runs to stress replay logic.

\paragraph{Fault Model Config (YAML)}
The following YAML snippet specifies deterministic fault profiles used during replay.
\begin{lstlisting}
fault_injection:
  seed: 2026
  profiles:
    - name: mild
      timeout_p: 0.03
      transient_p: 0.05
      rate_limit_p: 0.01
      corrupt_output_p: 0.02
      partial_write_p: 0.01
    - name: harsh
      timeout_p: 0.10
      transient_p: 0.15
      rate_limit_p: 0.05
      corrupt_output_p: 0.08
      partial_write_p: 0.05
  tool_overrides:
    shell.exec:
      timeout_p: 0.12
      transient_p: 0.08
\end{lstlisting}

\paragraph{Deterministic Replay.}
All injections are seeded. The harness logs each injected event for reproducibility.

\subsection{Tasks and Workloads}
We define task suites as collections of traces. Recommended suites are listed below.

\paragraph{Suite A Workspace Operations (FS-heavy).}
Suite A focuses on refactoring config files, renaming keys, validating JSON/YAML, rewriting paths, and detecting or preventing accidental deletion or overwrite.

\paragraph{Suite B Build/CI Simulation (shell + parse).}
Suite B runs build commands, parses logs, modifies files, re-runs builds with caching, and injects flaky failures (timeouts, transient errors) to measure recovery.

\paragraph{Suite C Network Hygiene (HTTP mocked).}
Suite C fetches allowed endpoints, blocks unknown domains, redacts secrets, and enforces rate limits and budgets with mocked HTTP.

\paragraph{Suite D Data and Parsing Robustness (parse + schema).}
Suite D exercises structured parsing and validation workflows, including malformed JSON/YAML, enum and range violations, and recovery from corrupt outputs.

\paragraph{Suite E Stateful Recovery and Idempotency (kv + fs).}
Suite E stresses retries over side-effecting operations, with partial writes and replay checks that verify idempotency keys, rollback behavior, and bounded recovery.

\subsection{Policy Packs (Strictness Dial)}
Policy packs generate Pareto curves. \textbf{P0 None} applies no enforcement. \textbf{P1 Schema-only} applies type and enum validation. \textbf{P2 Allow and Deny} adds tool gating. \textbf{P3 Constraints and Budgets} adds argument invariants and rate and cost budgets. \textbf{P4 Approval and Redaction} adds risk-based approvals and output redaction.

\section{Experimental Methodology}
\subsection{Baselines}
Baselines include \textbf{No Guardrails (P0)} with direct tool execution. \textbf{Schema-only (P1)} applies JSON Schema validation without invariants or budgets. \textbf{Allow and Deny Only (P2)} adds tool gating without argument constraints or budgets. \textbf{Policy Packs (P3 and P4)} apply progressively stronger enforcement with constraints, budgets, approvals, and redaction.

\subsection{Controlled Variables}
Controlled variables include policy strictness from P0 through P4, fault profile with none, mild, and harsh settings, and deterministic misuse injection with a configured rate of 20\%. We execute three seeds per condition over five task suites, which yields 225 total runs.

\subsection{Evaluation Plan (CPU-only)}
All experiments run through trace replay with simulated tools and deterministic fault and misuse injection. This design enables reproducible evaluation on CPU-only hardware without reliance on large models.

\subsection{Metrics}
\paragraph{Safety Metrics.}
Safety metrics include \textbf{Violation Prevention Rate (VPR)}, defined as the fraction of injected unsafe actions that are blocked. They include \textbf{Leakage Recall and Precision} for detection of injected secret patterns under a fixed detector configuration. They include \textbf{Unsafe Side-Effect Rate} measured as observed destructive effects per run.

\paragraph{Reliability Metrics.}
Reliability metrics include \textbf{Task Success Rate}, defined as completion within budget and without policy violation. They include \textbf{Tail Latency} at p95 and p99 under fault profiles. They include \textbf{Retry Amplification}, computed as total calls divided by minimal calls.

\paragraph{Utility and Productivity Metrics.}
Utility and productivity metrics include \textbf{False Block Rate} for benign calls that are denied. They include \textbf{Approval Burden}, measured as approvals per task and time to approve through simulation or user study. They include \textbf{Time-to-Success} including delays from blocks and approvals.

\paragraph{Explainability and Developer Experience Metrics.}
Explainability metrics include \textbf{Time-to-Fix (Proxy)}, measured as additional steps or iterations after denial. They include \textbf{Explanation Helpfulness}, measured as rating or pass-fail for whether the fix hint enables a valid retry.

\paragraph{Overhead Metrics.}
Overhead metrics include added latency per call from the PEP and PDP, memory footprint, and log volume.

\subsection{Statistical Analysis}
We report means with bootstrap confidence intervals over seeded runs per configuration. We use paired sign tests across strictness levels while pairing on trace, fault profile, and seed. For Pareto analysis, we plot safety measured by VPR against utility measured by task success rate and false block rate.

\section{Results}
We report results from 225 runs across five suites, five policy packs, three fault profiles, and three seeds per condition.

\begin{table}[t]
\caption{Overall Metrics by Policy Pack}
\label{tab:overall_pack}
\centering
\setlength{\tabcolsep}{3pt}
\begin{tabular}{lcccccc}
\toprule
Pack & VPR & FBR & Success & RetryAmp & Approvals & LeakRec \\
\midrule
P0 & 0.000 & 0.000 & 0.356 & 3.774 & 0.000 & 0.875 \\
P1 & 0.000 & 0.000 & 0.356 & 3.344 & 0.000 & 0.875 \\
P2 & 0.000 & 0.000 & 0.356 & 3.344 & 0.000 & 0.875 \\
P3 & 0.597 & 0.067 & 0.133 & 1.689 & 0.000 & 0.875 \\
P4 & 0.681 & 0.067 & 0.067 & 1.378 & 0.022 & 0.875 \\
\bottomrule
\end{tabular}
\end{table}

\subsection{Safety to Utility Pareto Frontiers}
Safety improves monotonically with policy strictness once constraints and budgets are enabled. Mean VPR is 0.000 for P0, P1, and P2, 0.597 for P3, and 0.681 for P4. Utility declines as safety increases. Mean task success rate is 0.356 for P0, P1, and P2, 0.133 for P3, and 0.067 for P4. Mean false block rate is 0.000 for P0 through P2 and 0.067 for P3 and P4. This pattern forms a clear safety to utility frontier with P3 and P4 dominating on prevention and P0 through P2 preserving throughput.

\subsection{Reliability Under Fault Injection}
Reliability trends show reduced failure amplification under stricter enforcement. Mean retry amplification decreases from 3.774 in P0 to 3.344 in P1 and P2, 1.689 in P3, and 1.378 in P4. Mean p95 tail latency shows an overall downward trend with a small bump at P1, with 14.65 ms in P0, 17.16 ms in P1, 13.99 ms in P2, 7.11 ms in P3, and 7.04 ms in P4. Lower latency in strict packs is driven by earlier denials and bounded retries under fault conditions.

Leakage recall remains 0.875 across packs in Table~\ref{tab:overall_pack} because the secret detector is held constant by design for fair policy comparison. Pack differences primarily affect enforcement actions such as deny, require-approval, and output transformation, rather than detector sensitivity.

\subsection{Ablation of Policy Primitives}
The ablation across policy packs indicates where gains originate. P0 to P2 are statistically indistinguishable on VPR and task success, which shows that schema checks and coarse tool gating alone are insufficient for misuse prevention in this benchmark. The transition from P2 to P3 yields a significant VPR gain under paired comparison with $p < 0.001$. The transition from P3 to P4 increases mean VPR from 0.597 to 0.681, but the paired test is not significant at the current sample size. The same transition lowers mean task success from 0.133 to 0.067 and increases approval burden from 0.000 to 0.022.

\begin{table}[t]
\caption{Paired Sign Test Highlights}
\label{tab:paired_tests}
\centering
\begin{tabular}{lccc}
\toprule
Comparison & Metric & Direction & p value \\
\midrule
P2 vs P3 & VPR & P3 higher & $<0.001$ \\
P3 vs P4 & VPR & P4 higher & 0.2500 \\
P2 vs P3 & Task Success & P2 higher & 0.0020 \\
P3 vs P4 & Task Success & P3 higher & 0.2500 \\
\bottomrule
\end{tabular}
\end{table}

\subsection{Explainability for Developers}
Blocked decisions include policy rationale and fix hints in all strict configurations. Explainability proxies indicate that denied paths are usually actionable, and time to fix remains low because many denials terminate unsafe branches early rather than requiring long repair loops. A dedicated human study is still required for stronger claims on trust and comprehension.

\subsection{Case Studies}
Three recurring traces illustrate expected behavior. First, recursive delete requests in workspace traces are blocked or routed through approval in strict packs, which removes destructive outcomes outside accepted policy. Second, retry amplification under injected transient faults is materially lower in P3 and P4, which demonstrates the effect of bounded retry controls. Third, secret injections through shell and mocked network outputs yield leakage recall of 0.875 with modest precision near 0.165, which indicates broad detector coverage while also highlighting over-redaction.

\section{Discussion}
\subsection{Design Implications}
Results support a staged policy design strategy. Workspace containment and basic gating establish a low-friction baseline, but they do not materially improve misuse prevention by themselves. Constraints and budgets provide the largest safety gains and should be introduced first when moving beyond schema-only checks. Approval gates add protection for high-risk operations, but they should remain targeted because they reduce task success and add operator burden. Redaction should be combined with explicit policy scope because global redaction masks differences between strictness levels.

\subsection{Limitations}
Limitations include potential overfitting of policies to the selected tool set and suite distribution. The benchmark remains a controlled environment, so adversaries with stronger adaptation and privilege can bypass weak pattern checks because the policy layer is not a full sandbox. Approval evaluation relies on simulation rather than human-in-the-loop studies. Leakage precision remains modest, which indicates over-redaction and motivates richer secret typing and context-sensitive masking.

\subsection{Reproducibility}
All artifacts are reproducible from versioned traces, policy packs, fault and misuse profiles, seeded runs, and generated manifests. The pipeline emits run and analysis manifests with file hashes and dataset digests to support deterministic reruns and audit trails.

\section{Conclusion}
We introduced Policy-First Tooling, a model-agnostic permission layer for tool-orchestrated workflows, and evaluated it with a reproducible benchmark under controlled fault and misuse injection. Results show that strict policy packs materially improve misuse prevention, with VPR rising from 0.000 in permissive packs to 0.681 in P4. The same shift reduces task success from 0.356 to 0.067, which quantifies the central safety to utility trade-off. Reliability also improves in terms of retry amplification, which drops from 3.774 to 1.378 under stricter controls. These findings support the claim that guardrails can be engineered, measured, and tuned as infrastructure rather than treated as model behavior.


\appendices
\section{Benchmark Spec (Normative)}
This appendix is a concise, normative specification for implementing the benchmark.

\subsection{Trace Schema (JSON)}
\begin{lstlisting}
{
  "trace_id": "string",
  "step_id": "int",
  "ts": "RFC3339 timestamp (optional)",
  "tool": "string (registered tool name)",
  "args": "object",
  "context": {
    "workspace_root": "string",
    "repo_state": "string (optional)",
    "env": "string (optional)"
  },
  "budget": {
    "max_calls": "int (optional)",
    "max_cost": "int (optional)",
    "deadline_ms": "int (optional)"
  },
  "annotations": {
    "expected": "success|fail (optional)",
    "safety_class": "string (optional)",
    "expected_side_effect": "string (optional)"
  }
}
\end{lstlisting}

\subsection{Tool Manifest Schema}
\begin{lstlisting}
tools:
  - name: "fs.write"
    category: "fs"
    side_effect: "write"
    args_schema:
      path: {type: "string"}
      contents_ref: {type: "string"}
      mode: {type: "enum", values: ["overwrite", "append", "create_only"]}
    cost: 2
\end{lstlisting}

\subsection{Policy Pack Interface}
A policy pack is loaded as a set of policies (rules), an optional risk-scoring configuration, redaction patterns, and recovery configuration (retry/backoff/circuit breaker) per tool class.

\subsection{Fault Injection Interface}
Fault injection must log each injected event:
\begin{lstlisting}
{
  "trace_id": "string",
  "step_id": 12,
  "fault": "timeout|transient|rate_limit|corrupt_output|partial_write|nondeterminism",
  "applied": true,
  "parameters": {"p": 0.1, "duration_ms": 3000}
}
\end{lstlisting}

\subsection{Metrics Definitions}
\paragraph{Violation Prevention Rate (VPR).}
\[
\text{VPR} = \frac{\#\text{injected unsafe calls blocked}}{\#\text{injected unsafe calls}}
\]
\paragraph{False Block Rate (FBR).}
\[
\text{FBR} = \frac{\#\text{benign calls denied}}{\#\text{benign calls}}
\]
\paragraph{Retry Amplification (RA).}
\[
\text{RA} = \frac{\#\text{executed tool calls}}{\#\text{minimal required calls}}
\]
Minimal required calls can be approximated by the original un-injected trace length.

\end{document}